\documentclass[%
 aip,
cp,  
 amsmath,amssymb,
 reprint,%
]{revtex4-2}

\usepackage{wrapfig}
\usepackage{sidecap} 
\usepackage{graphicx}
\usepackage{dcolumn}
\usepackage{bm}
\usepackage[colorlinks=true,unicode]{hyperref}
\usepackage{subcaption}

\usepackage[utf8]{inputenc}
\usepackage[T1]{fontenc}
\usepackage{mathptmx} 

\begin{document}

\title{Response of a Liquid $^3$He Neutron Detector}

\author{A.E.~Sharbaugh} 
 \email[Corresponding author: ]{alya.sharbaugh@ucdenver.edu}
 \affiliation{Department of Physics, University of Colorado Denver, Denver, Colorado 80217, USA}
\author{L.~Jones}%
 \email{luke.jones@ucdenver.edu}
 \affiliation{Department of Physics, University of Colorado Denver, Denver, Colorado 80217, USA}

\author{A.N.~Villano}
 \email{anthony.villano@ucdenver.edu}
 \affiliation{Department of Physics, University of Colorado Denver, Denver, Colorado 80217, USA}

\date{\today} 

\begin{abstract}
The $^3$He(n,p) process is excellent for neutron detection between thermal and $\sim$4\,MeV
because of the high cross section and near-complete energy transfer from the neutron to the proton.
This process is typically used in gaseous forms with ionization readout detectors. Here we study
the response of a liquid $^3$He neutron detector with a scintillation readout. We anticipate an
efficiency boost of around a factor of 64 compared to 10\,atm gaseous detectors, given similar
detector volumes. 
\end{abstract}

\maketitle

%
%
%
%

%
%
%
%
%
%

\section{\label{sec:intro}Introduction}

The (n,p) reaction occurs when a neutron enters the nucleus of an atom, simultaneously causing a
proton to leave. This process provides a reliable method for measuring the neutron flux through a
particular volume, provided that the cross section for the process is high enough and the Q-value
is positive or not very negative. Furthermore, since in this process the outgoing proton carries
an energy deterministically related to the incoming neutron energy, it is one of the only
measurements that can measure \emph{incoming} neutron energies unambiguously. 

For our neutron detector design, we primarily utilize the reaction $^3$He(n,p)$^3$H whose Q-value
can be ascertained by subtracting the proton separation energy of $^4$He from the neutron
separation energy of the same. The separation energies are obtained from the most recent Atomic
Mass Evaluation (AME) in 2020~\cite{Huang_2021,Wang_2021}. The resulting Q-value for the reaction
$^3$He(n,p)$^3$H is 763.755\,keV. Our design differs from previous detectors using this process
because those detectors were $^3$He gas-based with typical fill pressures between 4--10\,atm
partial pressure of $^3$He and equipped with ionization
readouts~\cite{BEIMER1986402,LANGFORD201351,BEST20161}. Our design implements $^3$He in the liquid
state which is at least 64 times more dense~\cite{PhysRev.96.551} than the gas detectors mentioned
(assuming a maximum $^3$He partial pressure of 10\,atm).The design has a scintillation readout via
a traditional photomultiplier tube (PMT); silicon photomultipliers are frozen-out at the proposed
operating temperatures. 

\begin{figure}[h!]
    \includegraphics[width=0.6\linewidth]{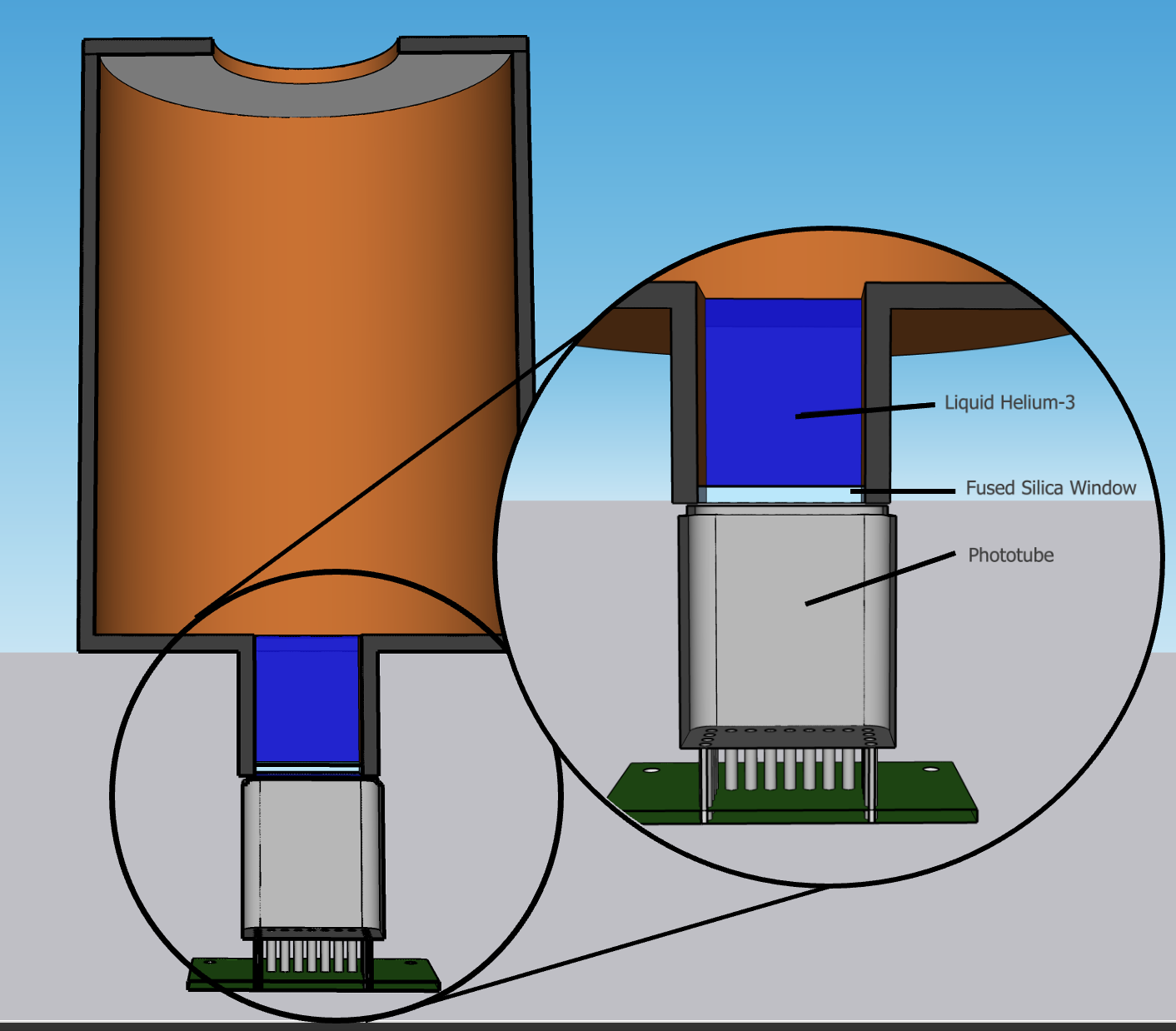}
    \caption[Detector Drawing]{The prototype design of a liquid $^3$He neutron detector with a
photomultiplier readout. The height of the gaseous portion of the vessel is approximately 11\,cm
tall, whereas the liquid fills a 2$\times$2$\times$2\,cm$^3$ cubical vessel attached near the bottom.
}
    \label{fig:detector-drawing}
\end{figure}

Recently the SPICE/HeRALD collaboration provided a proof-of-principle for a cryogenic
photomultiplier readout of a liquid helium volume~\cite{PhysRevD.105.092005}. The collaboration
also kindly shared their published data with us in digital form which we use for our design. Our
basic design is shown in Fig.~\ref{fig:detector-drawing}. We have envisioned a hermetic copper
vessel with a large gas space above a small cube (2$\times$2$\times$2\,cm$^3$) where liquid $^3$He
will collect when the vessel is filled with 6.63\,liters (22.9\,atm) of $^3$He at room temperature
and lowered to a temperature of 1\,K. Liquification will be achieved by attaching the vessel to an
off-the-shelf, commercially obtainable closed-cycle 1\,K cryocooler~\cite{cryocooler}. The cubical
portion of the vessel will have a small quartz window through coated with tetraphenyl butadiene
(TPB) so the phototube can measure the helium scintillation~\cite{MCKINSEY1997351}.

There are two key applications that we had in mind when designing this neutron detector:
measuring neutron backgrounds in the low-flux environments associated with rare event searches,
and measurements of neutron-producing nuclear reactions with tiny cross sections. Rare event
searches like dark matter and neutrinoless double-beta decay experiments are sensitive to neutron
backgrounds and often operate deep underground to shield out as many neutrons as possible. Still,
there are residual neutrons in the environment for various reasons~\cite{Dokania_2014}. The most
common method used to estimate the level of the residual neutrons is by simulation: either of the
($\alpha$,n) process for radiogenic neutrons~\cite{COOLEY2018110} or of cosmic rays for cosmogenic
neutrons~\cite{PhysRevD.73.053004}. One key difficulty is that, while the simulations themselves
have made great strides, there is often no measured spectrum to compare to because the neutron
flux is so low that its measurement comes with large effort~\cite{BEST20161,Browne:1999pe}. We
will focus the current study on the sensitivity of our detector design to neutrons in a deep
underground environment (SNOLAB). 

The measurements of neutron-producing nuclear reactions will not be discussed in detail but it is
worth mentioning this as another application for our detector design and potentially critical to
the field of nuclear astrophysics.  Nuclear astrophysicists have sought for decades to understand
the slow-neutron-capture process (s-process) in stars. The s-process is responsible for the
production of almost half the heavy elements in the universe. In order to constrain this process
hundreds of nuclear cross sections must be measured. One example of them is the
$^{13}$C($\alpha$,n)$^{16}$O reaction, which serves as a neutron source in stars near the Gamow
peak, between 150–230 keV.  Current measurements for that region yield uncertainties of greater
than 40\%~\cite{PhysRevLett.127.152701}. Another key example is the $^{22}$Ne($\alpha$,n)$^{25}$Mg
which competes with the similar ($\alpha$,$\gamma$) process that does not produce
neutrons~\cite{10.1140/epja/s10050-023-00917-9}. Each of those reactions have pico-barn level
cross sections in the off-resonance regions and could be measured precisely with an instrument
that has a high neutron detection efficiency. 
\section{\label{sec:spectral}Spectral features of $^3$He(n,p)}

Beimer, et al.~\cite{BEIMER1986402} outlines a procedure for computing the expected
deposited-energy spectra when a monoenergetic neutron impinges on a gaseous $^3$He detector
cylinder of 5\,cm diameter and 15\,cm length. We have reproduced the Beimer model for all incident
neutrons energies by interpolating their measured spectral parameters.  We are specifically
interested in the proton leakage portion of the distribution. For liquid helium detectors--like
the prototype we're designing--we expect the response to be qualitatively similar except with
substantially less proton leakage coming from the possibility of a proton escaping the sensitive
area of the detector after the $^3$He(n,p) reaction.

\begin{figure}[b!] \centering \includegraphics[width=1.0\linewidth]{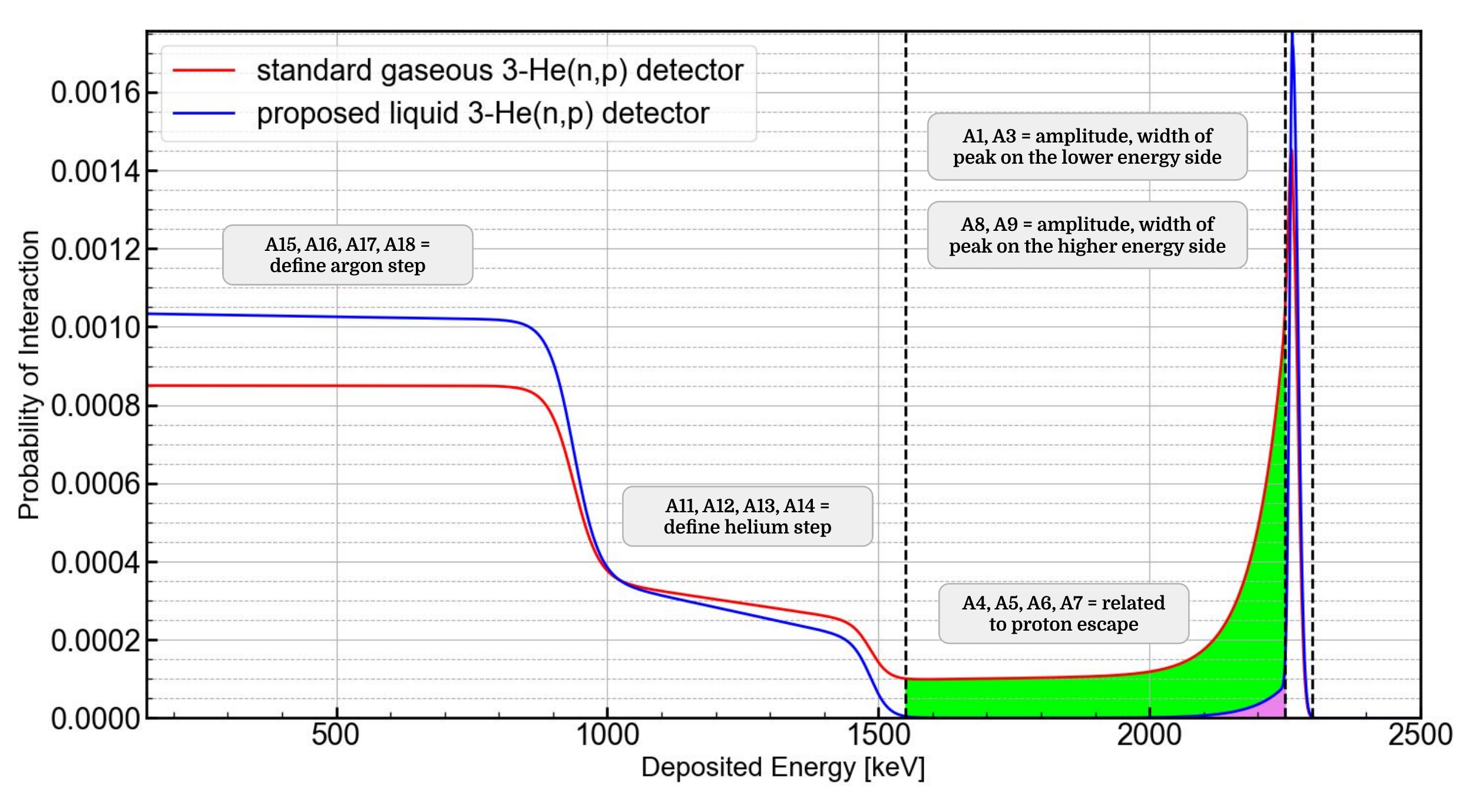}
   \caption[3He Response Function]{The parameters of the response function mapped to a 1.5\,MeV
signal peak. The red curve represents data collected by Biemer et al. using ionization vessels
filled with a gaseous mixture containing a partial pressure of 6\,atm of helium, 3\,atm of argon,
and 0.5\,atm of methane~\cite{BEIMER1986402}. Any scattering produced by the methane is
negligible. Parameters $A_{11}-A_{14}$ describe the scattering step formed by the argon, but this
part of the curve is not relevant to our analysis because the proposed detector will contain
purely helium. The blue curve represents the predicted spectra for neutrons of $E_n$=1.5\,MeV in
the liquid $^3$He detector. The area shaded green highlights the 80\% leakage from the standard
gaseous detector, while the area shaded purple highlights the 13.5\% leakage from the proposed
detector. Reducing proton escape increases peak efficiency, resulting in a taller and narrower
blue curve. 
}
    \label{fig:np_spectra}
\end{figure}

In the Beimer reference, the response of the $^3$He detector to neutrons was modeled by
construction a probability distribution function (PDF) composed of elements from each of the
physical processes involved. The processes were: elastic scattering from argon; elastic scattering
from helium; proton leakage; and the (n,p) peak. The original reference used 17 parameters ($A_1$
-- $A_{18}$; no $A_{10}$) to fit these key parts of the measured spectra to the response PDF, as
seen in Fig.~\ref{fig:np_spectra}. The important portions of the spectrum for our purposes are the
(n,p) peak, proton leakage, and elastic $^3$He contribution, as we expect these processes to also
be present in our liquid $^3$He prototype.

The (n,p) peak is governed by the terms in Eq.~\ref{eq:peak}. We define $f_p$ to be the PDF of
measured energy $E$ for a neutron with energy $E_n$. Amplitude, $A_1$ and $A_8$, behave similarly
but the width of the Gaussian is different on the lower energy side ($A_3$) than on the higher
energy side ($A_9$). $Q$=763.755\,keV represents the nuclear $Q$ value for the $^3$He (n,p)
reaction discussed above.

\begin{equation}
\label{eq:peak}
f_p(E, E_n) = \\
  \begin{cases}
  A_1e^{-\frac{1}{2}(\frac{(E_n+Q)-E}{A_3})^2} ; E \leq (E_n+Q) \\ 
  A_8e^{-\frac{1}{2}(\frac{(E_n+Q)-E}{A_9})^2} ; E > (E_n+Q)
  \end{cases} \\
\end{equation} \\

The elastic $^3$He contribution is governed by the terms in Eq.~\ref{elastic_3He}. This result
depends implicitly on the incoming neutron energy $E_n$ because the parameter $A_{13}$ scales with
this energy. The term $A_{13}-E$ gives a slightly negative slope at low energies. The remaining
parameters in the numerator ensure the PDF is never negative for a given energy below the maximum
recoil energy. 

\begin{equation}\label{elastic_3He}
f_{he}(E, E_n) = \frac{A_{11}(A_{13}-E)+A_{12}}{1+e^{\frac{E-A_{13}}{A_{14}}}}
\end{equation} \\

Parameters $A_4$ and $A_6$ describe the behavior of the exponential preceding the primary signal
peak. The population of this region, shaded green or purple in Fig.~\ref{fig:np_spectra}, is
determined by the number of protons which escape the active area and leak out of the detection
medium. We define $f_{esc}$ in Eq.~\ref{proton_escape}. Note parameters $A_2$ and $A_{10}$ are
unused~\cite{BEIMER1986402}. 

\begin{equation}\label{proton_escape}
f_{esc}(E, E_n) = \frac{1}{2}erfc(1-\frac{(E_n+Q)-E}{A_3})(
A_1A_4e^{-\frac{(E_n+Q)-E}{A_5}}+A_6e^{-\frac{(E_n+Q)-E}{A_7}})
\end{equation} \\

Based on our calculations, the leakage accounts for approximately 80\% of the original peak.
Protons (for 1.5\,MeV incident neutrons) in gaseous detectors of this type have a stopping length
of approximately 10\,cm. The gaseous $^3$He detectors used by Biemer et al. were cylindrical, with
a diameter of 5\,cm and a length of 15\,cm, which led to many protons escaping. By contrast,
protons in the proposed liquid $^3$He detector are estimated to have a stopping length of no more
than $\sim$0.1\,cm for up to around 2\,MeV neutron energies. The liquid $^3$He chamber will be a
cube with dimensions 2\,cm$\times$2\,cm$\times$2\,cm. As a result, we anticipate only minor escape
near the surfaces of the cube. Using the 0.1\,cm stopping length as a thickness, this surface
accounts for 27\% of the total volume of the detector. We can further estimate that half of these
interactions will rebound into the detector, resulting in an upper bound of 13.5\% leakage. This
assumption allows us to modify parameters $A_4$ and $A_6$ in our model. The complete probability
distribution function (PDF) obtained by summing each equation can be seen in
Fig.~\ref{fig:np_spectra}. 

A notable efficiency boost between a gaseous detectors of Beimer and our designed liquid detector
is expected. $^3$He has higher density in the liquid state than in the gaseous
state~\cite{PhysRev.96.551}. The density increase of the liquid $^3$He detector we propose is a
factor of 107, given a 6\,atm fill pressure of the gaseous tube we compare to. The actual
thickness of the gaseous detectors used by Beimer~\cite{BEIMER1986402} were 5\,cm thick, whereas
our design is 2\,cm thick; we therefore expect an efficiency gain of about 43$\times$ for neutrons
going straight through the 5\,cm dimension. This means that whereas Beimer~\cite{BEIMER1986402}
measured 0.03\% and 0.02\% peak efficiencies for their gaseous $^3$He detectors at 1\,MeV and
2.5\,MeV neutron energies respectively, we would expect efficiencies of 1.5\% and 0.88\%
respectively. For neutron energies of order 100\,keV the Beimer gaseous detectors have measured
efficiencies around 0.1\% so we would expect at least a 4.4\% efficiency. By understanding the
parameters governing the $^3$He(n,p) spectra, we can now extrapolate the model for cryogenic
($\sim$ 1\,K) temperatures and assess the performance of the proposed liquid $^3$He detector. 
\section{\label{sec:berkeley}Berkeley model of liquid $^4$He response}

Recently it has been demonstrated by the SPICE/HeRALD collaboration at Berkeley that liquid $^4$He
can be used as a scintillation detector with low-temperature photomultipliers
(PMTs)~\cite{PhysRevD.105.092005}. The Berkeley data show the response of liquid $^4$He
scintillation via PMT measurements at specific energies for electron recoil (ER) events (radiation
that produces a recoiling electron in the detector medium) and events composed of nuclear recoils
(NRs) of $^4$He atoms. Since our liquid $^3$He prototype will utilize the (n,p) process we are
primarily interested in proton recoil events whose behavior probably lies between the Berkeley
data for electron and nuclear recoils. Nevertheless, we use these measurements as a starting point
for the expected detector response in a $^3$He scintillating liquid--assuming qualitative
similarity to $^4$He in terms of scintillation yield and resolution.  

\begin{figure}[h]
    \includegraphics[scale=0.7]{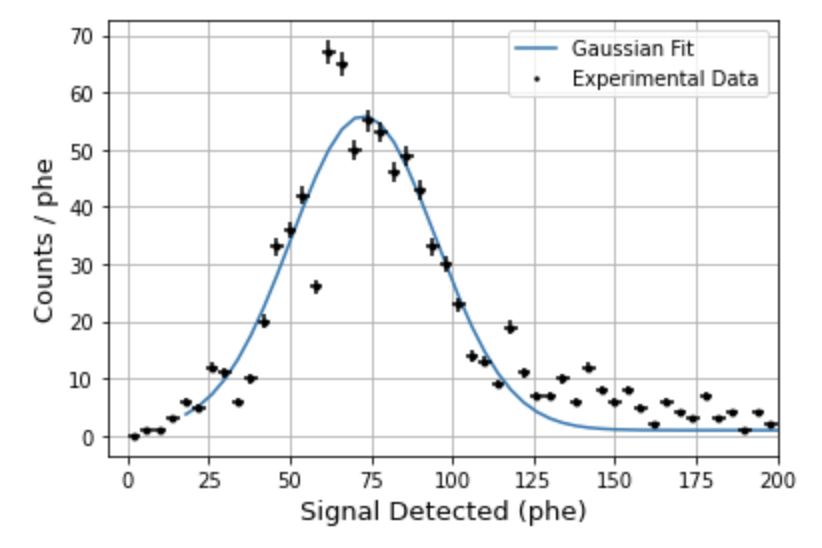}
    \caption{\label{fig:gpeak}The experimental data is example data taken
from~\cite{PhysRevD.105.092005}, corresponding to 142\,keV nuclear recoils. The horizontal
represents the signal size in number of photoelectrons (phe), and the vertical represents counts
per bin. 
}
\end{figure}

Our primary interest is to extrapolate this data, and predict the signal peak behavior of incident
protons within the range of 700\,keV - 3\,MeV. To do so, we need to define certain characteristics
of each signal peak. First, we fit a Gaussian curve to each signal peak in the Berkeley dataset
(see Fig.~\ref{fig:gpeak} for an example). In total, we make 13 Gaussian fits (6 ER, and 7 NR).
The fitting function is described by Equation \ref{eq:gauss}.

\begin{equation}\label{eq:gauss}
G(x) = a_1 \textrm{exp} \left(-\frac{1}{2}\left(\frac{x - a_2}{a_3}\right)^2\right) + a_4.
\end{equation}

\begin{figure}[h]
    \includegraphics[scale=0.5]{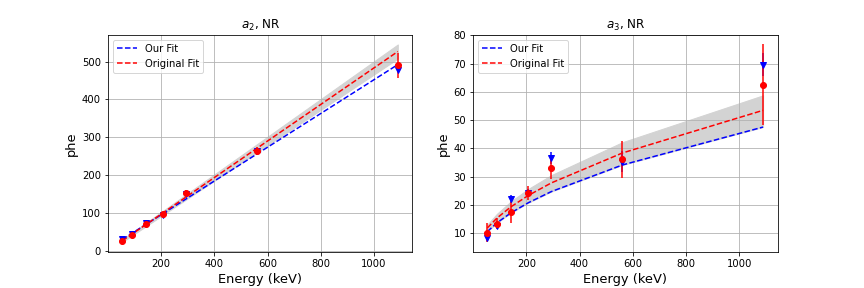}
    \vspace{0.2cm}
    \includegraphics[scale=0.5]{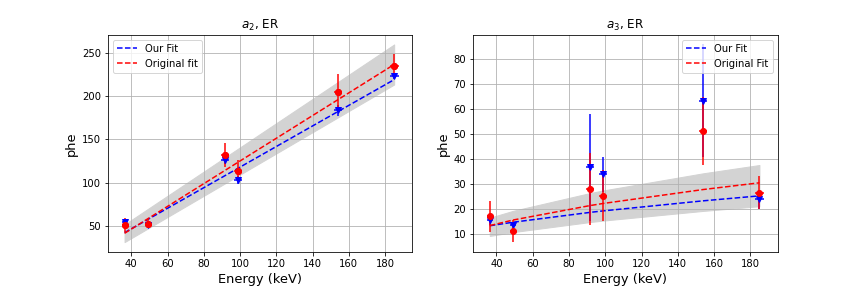}
    \caption{\label{fig:aifits}(Top) From left to right, the plots of $a_2$ and $a_3$ for nuclear
recoils, respectively. (Bottom) From left to right, the plots of $a_2$ and $a_3$ for electronic
recoils, respectively.  For each figure, the energy uncertainties are obtained
from~\cite{PhysRevD.105.092005} and the phe uncertainties are obtained from the covariance matrix
of the fits. The red points correspond to the individual photon yield/width data and uncertainties
from the Berkeley publication and the red dashed lines are fits to those points. The blue points
represent the yield/widths from our Gaussian fits and their uncertainties and the blue dashed
lines are a fit to those points. For each plot, we used the uncertainties from
Table~\ref{tab:fitresults} to obtain an upper and lower bound for the energy behavior, denoted by
the shaded regions centered around the red dashed line. Seeing as both the blue dashed line falls
within this region for each plot, we see that the fits are consistent.
}
\end{figure}

It should be noted that we obtain each data point from the same Gaussian curve with the
exception of the point corresponding to 1090 keV NR. Instead, we fit this point to the following
Gaussian plus linear curve denoted $G_0$:

\begin{equation}\label{eq:gaussplus}
 G_0(x) = a_1 \textrm{exp} \left(-\frac{1}{2}\left(\frac{x - a_2}{a_3}\right)^2\right) + a_4x + a_5,
\end{equation}

the reason for this is that the fit to the other function was not particularly good in this case,
so the new function allows for a more precise data point such that the uncertainty has a
consistent order of magnitude relative to all other data points. Throughout all of these fits, the
parameters we are interested in are $a_2$ and $a_3$, as these determine the peak signal size and
the width of our signal respectively, in number of photoelectrons (phe). After completing each
fit, we have two sets of $a_2$'s and two sets of $a_3$'s, one for nuclear and one for electronic
recoils. We then create a plot of each $a_i$ vs. recoil energy $E$. The fits of these plots are
outlined in Fig.~\ref{fig:aifits}.

For the $a_2$'s, we did a weighted fit to the collection of points to a linear function $y = mx +
b$, where $y$ is the signal size in npe, and $x$ is the event energy in keV. The results of these
linear fits using both our extracted yields and the Berkeley extracted yields are shown in
Fig.~\ref{fig:aifits}, left sides.

For the $a_3$ parameters we model the overall behavior in a similar way, except using a weighted
fit to a standard width-versus-energy function $y = \sqrt{Ax + \sigma_0^2}$. Again we do this fit
for both our extracted widths and the ones from the Berekely
publication~\cite{PhysRevD.105.092005} and the results are displayed in Fig.~\ref{fig:aifits}. It
was found that in all cases the $\sigma_0$ parameter was negligibly small, so we don't list it
with our results. The reason for this is that we don't have data at low enough scattering energies
to probe this baseline (zero-energy) resolution. After the fitting procedure, we end up with the
parameters in Table~\ref{tab:fitresults}.

\begin{table}[htp]

{\renewcommand{\arraystretch}{3}%
\begin{tabular}{c  c  c  c}
    \hline
    \hline 
	Recoil Type &  Slope   & Intercept & A  \\ \hline
        NR & $0.461 \pm 0.003$ & $4.310 \pm 0.556$ & $2.457 \pm 0.192$   \\ 
        ER & $1.185 \pm 0.023$ & $-0.338 \pm 1.610$ & $3.109 \pm 1.263$   \\ 
    \hline 
    \hline
	Recoil Type &  Slope   & Intercept & A  \\ \hline
        NR & $ 0.483 \pm 0.015$ & $-0.202 \pm 2.402$ & $2.621  \pm 0.548$  \\ 
        ER & $1.308 \pm 0.087$ & $-5.523 \pm 6.998$ & $5.052 \pm 2.613$ \\
    \hline
    \hline
\end{tabular}}
    
\caption{\label{tab:fitresults} (Top) A table of fit parameters which quantify how the Gaussian
parameters $a_2$ and $a_3$ behave with deposited energy according to our fits.  (Bottom) A table
of fit parameters which quantify how the Gaussian parameters $a_2$ and $a_3$ behave with deposited
energy according to the fits from~\cite{PhysRevD.105.092005}. The 'slope' and 'intercept'
parameters correspond to the slope and intercept of the $a_2$ plots. The parameter $A$ corresponds
to the non-linear fit function for the $a_3$ parameters. Again, we do not display values for the
parameter $\sigma_0$ because they are all near zero. These plots are shown in
Fig.~\ref{fig:aifits}}

\end{table}

The fit parameters in Table~\ref{tab:fitresults} show rough agreement with the analysis of the
Berkeley reference~\cite{PhysRevD.105.092005}. They likely differ because of the more precise fit
function for each scattering histogram and a more careful treatment of systematic errors. We will
use these models for the detector resolution in the next section despite the large extrapolation
necessary (the highest fitted energy is around 1\,MeV recoil for NRs and 180\,keV for ERs whereas
we need the detector performance at the $\sim$MeV scale for both NRs and ERs). 
\section{\label{sec:fullresponse}Liquid $^3$He wide-band neutron spectral response}


We expect our liquid $^3$He neutron detector design to be appropriate for low neutron flux
environments because of its high efficiency relative to conventional gaseous $^3$He tubes like
those used by Beimer~\cite{BEIMER1986402}. This is because of the much higher density of $^3$He in
the liquid state at 1\,K--0.0792\,g/cm$^3$--compared to a 6\,atm gas with density
7.4$\times$10$^{-4}$\,g/cm$^3$ at room temperature~\cite{PhysRev.96.551}. In fact, our
2$\times$2$\times$2\,cm$^3$ detector is expected register 3 times more neutron events above around
10\,keV than the larger tubes (5\,cm diameter; 15\,cm length) of Beimer~\cite{BEIMER1986402}. For
typical gaseous tubes with fill pressures between 6--10\,atm our design will be 64--107 times more
efficient per volume at neutron energies larger than 10\,keV. Below 10\,keV the efficiency will
quickly approach 100\%.  

\begin{figure}[h]
    \centering
    \includegraphics[width=1.0\linewidth]{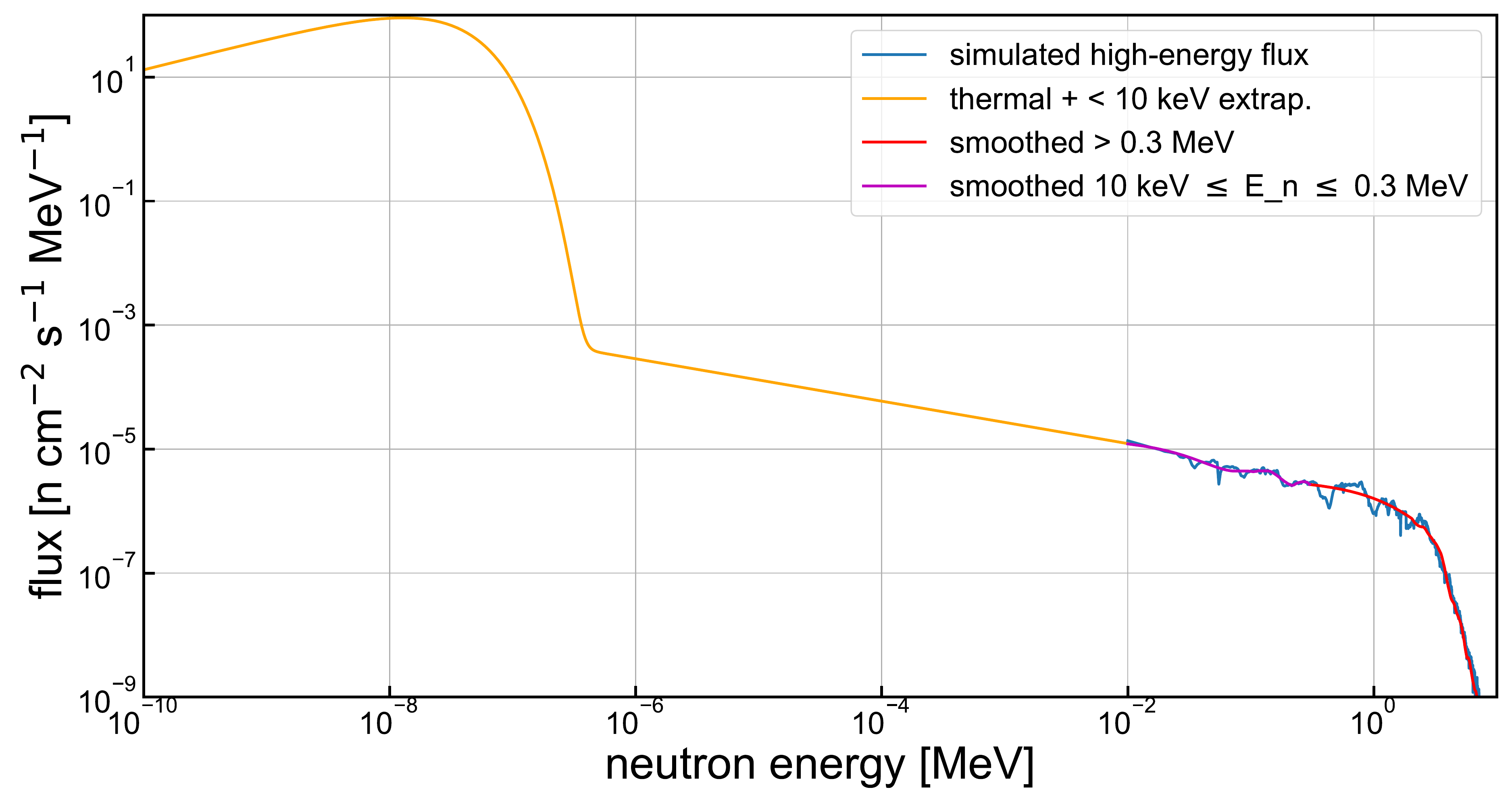}
    \caption[SNOLAB Flux]{Neutron flux as a function of energy for the SNOLAB environment. The red
and purple curves are smoothings of the high-energy neutrons computed from the SuperCDMS
sensitivity projection paper~\cite{PhysRevD.95.082002}. The orange portion of the curve is an
interpolation in the region below 10\,keV down to the Maxwell-Boltzmann portion for the thermal
neutron flux. The integral of the thermal flux region (from 10$^{-4}$\,eV to about 1\,eV) is
normalized to the measured underground SNOLAB thermal neutron flux: 4144.8$\pm$49.8$\pm$105.3
n/m$^2$/day.  
}
    \label{fig:flux_function}
\end{figure}

SNOLAB is home to some of the most sensitive low-background experiments and as such has an
exceptionally low neutron flux. We have estimated the neutron flux as a function of energy from
around 6--10\,MeV neutrons all the way down to thermal energies. These results are seen in
Fig.~\ref{fig:flux_function}. The high-energy flux is derived from the simulated neutron flux of
the Super Cryogenic Dark Matter Search (SuperCDMS) for the SNOLAB
environment~\cite{PhysRevD.95.082002}. In that flux function there were several sharp features due
to the modeling from SOURCES4C~\cite{WilsonJan2002} and the propagation of ($\alpha$,n) neutrons
through the cavern rock and shotcrete. We have smoothed over features (see
Fig.~\ref{fig:flux_function} red and purple curves) to produce a smooth flux function down to
around 10\,keV neutron energy. The normalization for the high-energy flux portion of this curve is
taken from the SNOLAB handbook~\cite{SNOLABhandbook} where the value is 4000\,n/m$^2$/day. The
high-energy flux is quoted as ``fast'' in the reference, this typically means above 1\,MeV but no
explicit lower bound is given. Below about 10\,keV in neutron energy we have extrapolated the flux
with a simple power-law function $aE^b$. The power law was determined by a linear regression of
the log of our smoothed data in the range 10--20\,keV. The extrapolated flux was joined at low
energies with a Maxwell-Boltzmann distribution for (thermal) neutrons at room temperature. The
unit-normalized function for the thermal neutrons is:

\begin{equation}
f_t(E) = 2 \sqrt{\frac{E}{\pi}}\left ( \frac{1}{kT} \right)^{3/2} exp \left ( - \frac{E}{kT}
\right ).
\end{equation} 
This thermal flux was multiplied by the measured total flux of thermal neutrons in
SNOLAB~\cite{Browne:1999pe}, 4144.9$\pm$49.8$\pm$105.3 n/m$^2$/day. In this quoted measurement the
first set of uncertainties is statistical and the second set is \emph{systematic}. The systematic
uncertainties come from inexactness in the analysis methods whereas the statistical uncertainties
are based on the limited number of counts observed over the finite measurement period.  

If the neutron flux as a function of energy from Fig.~\ref{fig:flux_function} is described as
$f(E)$, then we might expect the overall rate in our detector to be $R =
f(E)(\Sigma_{n,p}+\Sigma_e)V$, where $R$ is the differential rate at a given neutron energy;
$\Sigma_{n,p}$ is the macroscopic (n,p) cross section for $^3$He as a function of neutron energy;
$\Sigma_e$ is the elastic cross section for $^3$He as a function of neutron energy; and $V$ is the
detector volume. In fact this expression is only accurate for neutron energies above approximately
1\,keV. Below that value the cross section for the (n,p) interaction is so large that there is
significant \emph{self-shielding}. Self-shielding occurs when a material has a high enough
interaction probability that the neutrons will rarely penetrate to the center of the material.  In
the case of thermal neutrons (with energies around 0.025\,eV) the situation is extreme and all
(n,p) events will take place at the surface of the detector. 

\begin{figure}[h]
    \centering
    \includegraphics[width=1.0\linewidth]{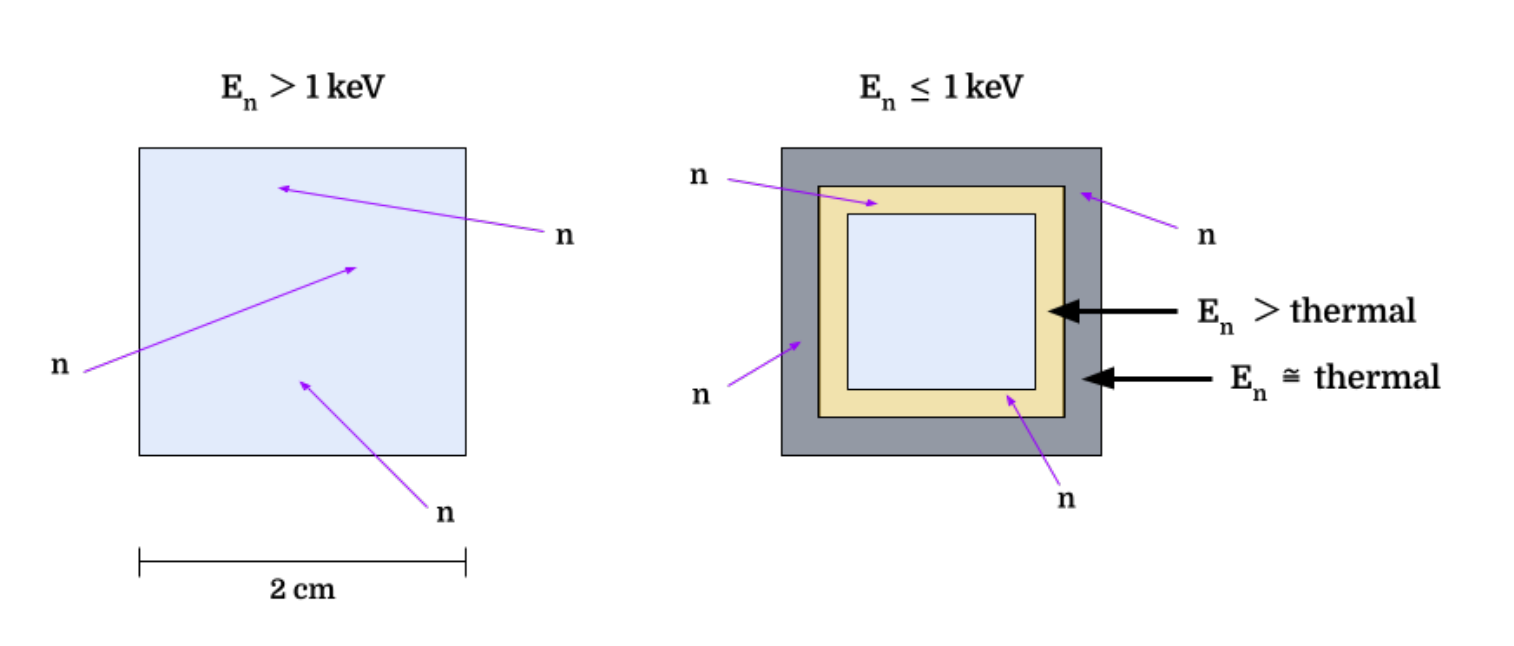}
    \caption[Flux Calc]{Schematic depicting how the flux is converted into an expected event rate.
(Left) Shows situations where the (n,p) cross section is small enough (mean free path of neutrons
much larger than detector size) that the flux can be treated as uniform throughout the detector
and equal to the external neutron flux. (Right) Shows situations--taken to be $E_n$<1\,keV--where
any incoming neutron is nearly guaranteed to produce an (n,p) reaction in the detector. The black
region near the edge depicts the region where thermal neutrons ($E_n \sim$0.025\,eV) are likely to
interact. The tan region shows where neutrons of energies somewhat higher--up to around
1\,keV--are likely to interact. 
    }
    \label{fig:self_shielding}
\end{figure}

We modeled the neutron flux below 1\,keV at the surface of our detector as non-zero only in the
direction toward the inside of the detector in a 2$\pi$ solid angle. This was accomplished by
using the following \emph{angular neutron flux} (following the notation of Duderstadt and
Hamilton~\cite{duderstadt1976nuclear}):

\begin{equation}
\varphi(\mathbf{r},E,\hat{\Omega},t) = \frac{1}{4\pi} f(E) \theta(-\hat{\mathbf{n}} \cdot
\hat{\Omega}),
\end{equation}
where $\hat{\mathbf{n}}$ is the unit normal of the surface of our detector and $\theta$ is the
unit step function. $\hat{\Omega}$ is the unit vector about point $\mathbf{r}$ in which the flux
is moving. The \emph{angular current density} is then given by the following expression.

\begin{equation}
\mathbf{j}(\mathbf{r},E,\hat{\Omega},t) = \hat{\Omega}\varphi(\mathbf{r},E,\hat{\Omega},t)
\end{equation}
Correspondingly, the rate of neutrons passing into our detector volume is:
\begin{equation}\label{eq:surfrate}
R_{\mathrm{surf}} = \oint \mathbf{j}(\mathbf{r},E,\hat{\Omega},t)\cdot dA dE d\hat{\Omega}.
\end{equation}
Below 1\,keV we have assumed that the detector is completely opaque and therefore the number of
neutron detections is given by Eq.~\ref{eq:surfrate} independent of the macroscopic cross section
$\Sigma_{n,p}$. This situation is depicted in Fig.~\ref{fig:self_shielding}. The left side shows
how the flux is treated above 1\,keV neutron energy--the detector is effectively transparent and
the expression $f\Sigma~V$ is an excellent approximation for the rate of detections. The right side
of Fig.~\ref{fig:self_shielding} shows that below 1\,keV a neutron entering the detector is highly
likely to interact near the surface, either right at the surface if the energy is very low (like
thermal neutrons) or slightly further in. To assist the reader in understanding the transition
between low neutron mean free path and high mean free path, we have included a plot of the mean
free path for neutrons in liquid and gaseous (6\,atm partial pressure) $^3$He in
Fig.~\ref{fig:mean-free-path}a.

The (n,p) interactions which do take place near the edge of the detector--mostly, but not all below
1\,keV--have a chance to have some of the proton energy escape the liquid. For gaseous detectors
(as discussed earlier) this possibility could account for large percentages (up to 80\%) of the measured
events. Because of the higher densities in liquid this possibility is highly suppressed but can
still happen. We computed, using the stopping power for protons in helium (obtained from the
\texttt{PSTAR} online database version \texttt{2.0.1},July 2017~\cite{PSTAR}), that 0.7\% of
low-energy events will have protons that ``leak'' out of the sensitive volume. We account for
these in our model by selecting randomly that percentage of events in a given energy region and
decreasing their deposited energy based on what percent of their stopping path is contained in the
active region. To give the reader a sense for when leakage will happen we included the range of
proton and alpha particles in $^3$He liquid and gas (at 6\,atm pressure) in
Fig.~\ref{fig:mean-free-path}b. 

For elastic scatters of neutrons on the $^3$He nuclei, the detector is always essentially
transparent because the macroscopic elastic cross section $\Sigma_{\mathrm{el}}$ is small.
Therefore elastic scattering does not modify the neutron flux throughout the detector and those
scatters are modeled in the same way as our (n,p) interactions at high-energy (above 1\,keV).

\begin{figure*}[h]
    \centering
    \begin{subfigure}[t]{0.5\textwidth}
        \centering
        \includegraphics[height=2.2in]{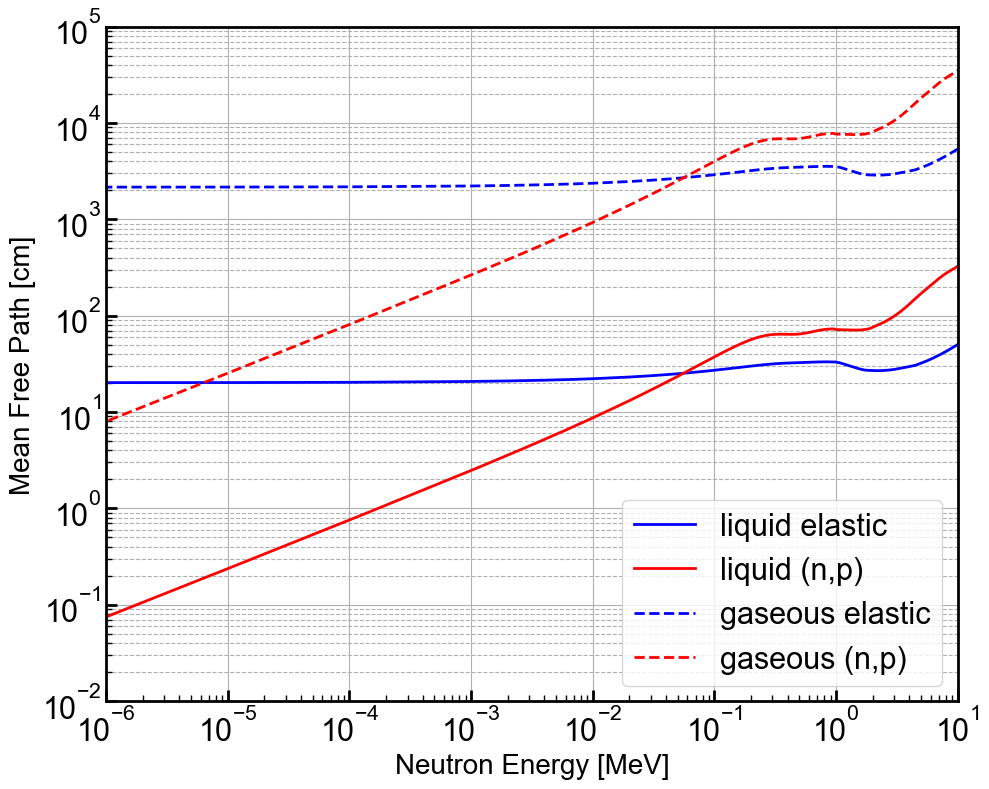}
        \caption{}
    \end{subfigure}%
    ~ 
    \begin{subfigure}[t]{0.5\textwidth}
        \centering
        \includegraphics[height=2.2in]{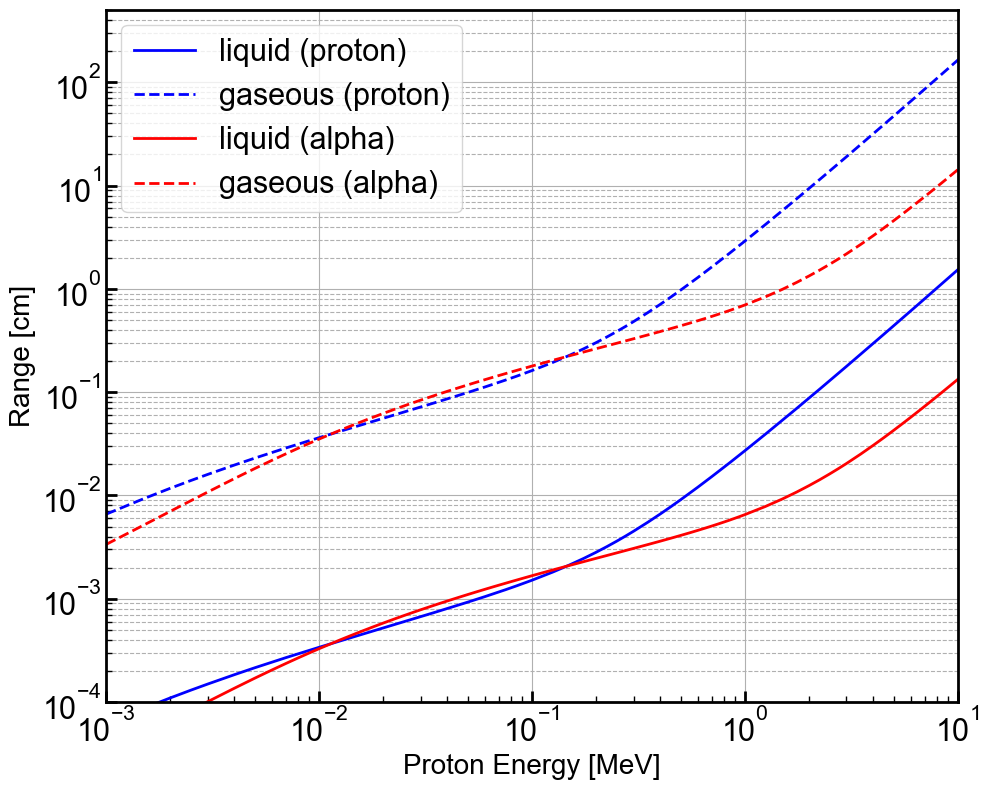}
        \caption{}
    \end{subfigure}
    \caption{The properties of relevant particles as they travel in gaseous and liquid $^3$He. (a)
The mean free path of neutrons in $^3$He for (n,p). Some neutrons will elastically scatter while
others will react via the (n,p) process. (b) The range of protons and alpha particles in $^3$He
which help determine the area of an energy event after a neutron interaction.}
    \label{fig:mean_free_path}
\end{figure*}

For both the (n,p) and elastic processes we use the evaluated cross sections as a function of
energy from the JENDL 5.0 compilation~\cite{jendl5_paper}. From those, we calculate the needed
macroscopic cross sections by scaling the cross section by the number density of the liquid
$^3$He. The behavior of the cross sections is reflected in Fig.~\ref{fig:mean_free_path}a. 

\begin{figure}[hbp]
    \centering
    \includegraphics[width=1.0\linewidth]{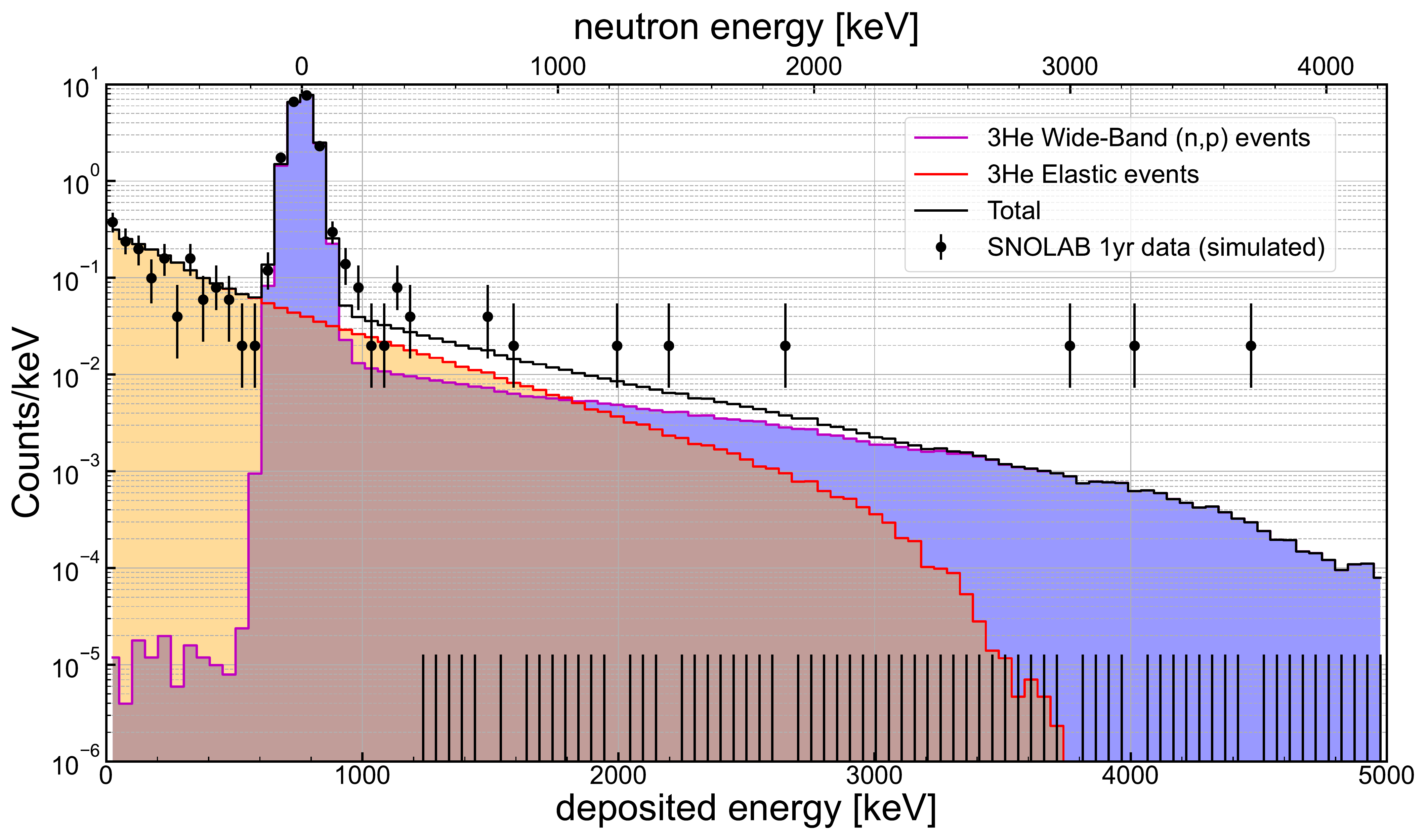}
    \caption[Expected Spectrum]{The expected spectrum of a 2$\times$2$\times$2\,cm$^3$ detector
operated at SNOLAB for 1\,yr (black points) compared to the expected total neutron distribution
(black line) with (n,p) (blue shaded) and elastic (orange shaded) contributions. The uncertainties
for zero-count bins were scaled so they did not obscure the plot, the points appear as black
spikes on the plot ending at a vertical axis value of 10$^{-5}$.  }
    \label{fig:response_function}
\end{figure}

Figure~\ref{fig:response_function} shows the expected distribution of events for a
2$\times$2$\times$2\,cm$^3$ detector of liquid $^3$He given the spectral model, anticipated
scintillation resolutions, cross-section weighting, and leakage corrections previously discussed.
We have used an NR detector response model for recoiling $^3$He and an ER response model for
recoiling protons. This treatment is crude, but plausible: modeling $^3$He as NRs implicitly
includes nuclear and electronic stopping powers, and the neglect of nuclear stopping power for
protons is a good approximation for the initial recoil energies we are interested in (above around
764\,keV). Our proton model is supported by Bethe-Bloch theory~\cite{ParticleDataGroup:2020ssz}
where electronic stopping is dominant and dependent on the relativistic velocity $\beta$ only.
Most of our proton stopping happens between $\beta=0.01$ and $\beta=0.04$ ($\sim$764\,keV)--within
the range where electron-stopping Bethe-Bloch theory is adequate for the purposes of a rough
(10--20\% accurate) analysis of our design.   

The figure shows that for a 1\,yr exposure we will produce an excellent measurement of the thermal
neutron flux--given by the size of the peak near a deposited energy equal to the Q-value of the
(n,p) reaction. It is also likely that a run period of that time will produce 3--4\,counts in the
high-energy neutron region above 2700\,keV. Neutrons of those energies will constrain the overall
fast flux present at the detector location. Similarly, the counts in the intermediate region,
above around 266\,keV neutron energy but below 1000\,keV will create an important constraint in a
region that has previously been difficult to measure.  
\section{\label{sec:conclusions}Conclusions}

While gaseous $^3$He tubes are probably better for thermal neutron measurements due to their high
surface area, our design is appropriate for cryogenic environments and above 1\,keV in neutron
energy our design surpasses gaseous tube performance.  A single liquid $^3$He detector of
dimensions 2$\times$2$\times$2\,cm$^3$ has superior efficiency to a gaseous $^3$He detector of the
same dimensions for $\sim$1\,keV--4\,MeV neutrons. Our current calculations also indicate that in
a reasonable running period (1\,yr) it can make impactful measurements in some of the lowest
neutron flux environments in the world, like SNOLAB. Based on previous measurements at
Berkeley~\cite{PhysRevD.105.092005} the resolution is expected to be useful for either low-flux
measurements or nuclear physics beam measurements of neutrons. 

Even though gaseous detectors are not very limited in size, scaling them will increase the
radiogenic alpha background proportionally to the surface area. The Helium and Lead Observatory
(HALO)--a gaseous $^3$He array operating at SNOLAB--is a good comparison~\cite{Bruulsema:2017}.
The array consists of 128 gaseous $^3$He detectors of length 3\,m or 2.5\,m and 5\,cm diameter at
a partial pressure of 2.5\,atm. Alpha backgrounds in this array are probably as low as can be
expected for gas detectors, using a special ultra-pure nickel tubes that reach a thorium
contamination of 1\,ppt~\cite{BOGER2000172}. The large size of the array means while our prototype
detector expects about 30 neutron events above 1\,keV in one year, the HALO array will expect
approximately 11,000.  However, the alpha background is far lower in our liquid $^3$He prototype
because of the dramatically lower surface area and the possibility of ultra-pure copper. The
thorium contamination has been demonstrated in copper as low as 0.01--0.03\,ppt by the Majorana
Demonstrator neutrinoless double-beta decay experiment~\cite{ABGRALL201622}. The Enriched Xenon
Observatory (EXO) has also demonstrated a contamination of thorium at 1.8\,ppt for Quartz--the
same material we will use for the PMT window~\cite{LEONARD2008490}.  Given those levels, we expect
a signa-to-background ratio of 0.079 in HALO but between 4.62 and 15.42 in our small
2$\times$2$\times$2\,cm$^3$ prototype, depending on the copper purity (between 0.01\,ppt and
1\,ppt). 

Our work indicates that this topic is worth further study. In particular, ambient gamma
backgrounds must be considered as well as those that might originate from the instrumentation of
our design--like from the photomultiplier tubes or electronics.  

\clearpage

\begin{acknowledgments}
The authors would like to thank Junsong Lin, A.J. Biffl, James deBoer, and Matt Pyle for helpful discussions.
We also gratefully acknowledge the SPICE/HeRALD collaboration for providing us with the data from
several plots in their paper. 
\end{acknowledgments}

%
%
\end{document}